\documentclass[runningheads]{llncs}

\usepackage[ruled, vlined, linesnumbered]{algorithm2e}

\usepackage{cmap}

\usepackage[ngerman,english]{babel}
\addto\extrasenglish{\languageshorthands{ngerman}\useshorthands{"}}

\usepackage{ifluatex}
\ifluatex
  \usepackage{fontspec}
  \usepackage[english]{selnolig}
\fi

\iftrue 
  \ifluatex
  \else
    \usepackage[%
      rm={oldstyle=false,proportional=true},%
      sf={oldstyle=false,proportional=true},%
      tt={oldstyle=false,proportional=true,variable=false},%
      qt=false%
    ]{cfr-lm}
  \fi
\else
  \ifluatex
  \else
    \usepackage{newtxtext}
    \usepackage{newtxmath}
    \usepackage[zerostyle=b,scaled=.9]{newtxtt}
  \fi
\fi

\ifluatex
\else
  \usepackage[T1]{fontenc}
  \usepackage[utf8]{inputenc} 
\fi

\usepackage{graphicx}

\usepackage{textcmds}

\usepackage{url}

\usepackage[absolute]{textpos}

\begin{document}
\title{ProCal: A Low-Cost and Programmable \\Calibration Tool for IoT Devices}
%
\titlerunning{ProCal: A Low-Cost and Programmable Calibration Tool for IoT Devices}  
%
\author{Chia-Chi Li\inst{1,2}\orcidID{0000-0003-2881-6354} \and Behnam Dezfouli\inst{1}\orcidID{0000-0001-6090-0412}}
\authorrunning{Li and Dezfouli} 
%
\institute{Internet of Things Research Lab, Department of Computer Engineering, \\Santa Clara University, California, USA\\
 \and Intel Corporation, Santa Clara, California, USA\\
\email{\{cli1, bdezfouli\}@scu.edu}
}

\maketitle              
\begin{abstract}
Calibration is an important step towards building reliable IoT systems.
For example, accurate sensor reading requires ADC calibration, and power monitoring chips must be calibrated before being used for measuring the energy consumption of IoT devices.
In this paper, we present ProCal, a low-cost, accurate, and scalable power calibration tool.
ProCal is a programmable platform which provides dynamic voltage and current output for calibration.
The basic idea is to use a digital potentiometer connected to a parallel resistor network controlled through digital switches.
The resistance and output frequency of ProCal is controlled by a software communicating with the board through the SPI interface.
Our design provides a simple synchronization mechanism which prevents the need for accurate time synchronization.  
We present mathematical modeling and validation of the tool by incorporating the concept of Fibonacci sequence.
Our extensive experimental studies show that this tool can significantly improve measurement accuracy.
For example, for ATMega2560, the ADC error reduces from 0.2\% to 0.01\%. 
ProCal not only costs less than 2\% of the current commercial solutions, it is also highly accurate by being able to provide extensive range of current and voltage values.

\end{abstract}

\section{Introduction} 
\label{intro}


One of the main capabilities of IoT devices is sensing. 
To this end, sensors such as temperature, pressure, humidity, and acoustic, are connected to analog-to-digital converters (ADC) of IoT devices.
Confidence about accuracy is particularly important for mission-critical applications such as monitoring a tank's temperature, a pipe's pressure in a chemical factory, or human vital signs~\cite{dezfouli2017rewimo}.
Without calibration, for example, a 10-bit ADC might result in a 5\textdegree{C} measurement error.
In addition, as IoT devices are usually battery powered, the energy of these devices should be carefully measured and optimized based on application requirements.
Studies show that it is important for the embedded power monitor to continuously keep track of power consumption, so that node failures can be diagnosed and eliminated~\cite{szewczyk2004analysis}.
The existing power monitoring devices make this possible by including ADCs that measure both voltage and current~\cite{dezfouli2018empiot,zhou2013nemo,lim2013flocklab,potsch2017efficient,jiang2007micro,haratcherev2008powerbench,hartung2016distributed}.
Consequently, it is essential to ensure that both sensor reading and power measurement are performed with high accuracy.
Multiple factors such as the inaccuracy of the ADC used, temperature, the supply voltage of the sensor, and the length of the sensor-to-ADC path can contribute to measurement inaccuracy~\cite{bychkovskiy2003collaborative}.
For example, the measurement error of INA219 power monitoring chip increases as its supply voltage reduces~\cite{dezfouli2018empiot}.

From the ADC point of view, measurement accuracy is influenced by its non-linearity.
Two of the key static specifications that define the accuracy of an ADC are \textit{differential non-linearity} (DNL) and \textit{integral non-linearity }(INL).  
DNL describes deviations from the ideal transition voltage in the converter's transfer functions.
The deviation from ideal transition is the differential linearity error for that unique code, which translates to \textit{least significant bit} (LSB).  
INL is the overall shape of the transfer function of ADC.  
This error is also referred to as \textit{static linearity }or \textit{absolute linearity}. 
The maximum deviation from the ideal transfer function to INL line is the worst case integral non-linearity~\cite{adc2015}. 
Non-linearity causes data distortion while the signals are being digitized. 
By thorough characterization of an ADC's non-linearity and transfer function, we can minimize its data distortion.
There are several different approximations of ADC non-linearity commonly used for calibration; such as common polynomials, Chebyshev polynomials, and Fourier series~\cite{suchanek2009adc}.
ADC calibration requires characterization of both increasing and decreasing input levels, which requires a wide range of samples~\cite{ieeeieee}. 
Since all components, including ADCs, in an analog chain demonstrate deviations from their normal value, a full calibration process is essential.

The two common approaches for calibrating an analog input are: (i) using different resistors~\cite{jiang2007micro,milenkovic2005environment}, and (ii) a potentiometer~\cite{zhou2013nemo,haratcherev2008powerbench} to generate variable loads.  
Although the former approach provides an accurate and stable load, it requires a large number of fixed-value resistors and a lot of effort to collect enough samples for a quality calibration.  
Due to these challenges, often times, researchers use only a few resistors to calibrate the full range of an ADC~\cite{milenkovic2005environment,jiang2007micro}.  
This insufficient use of resistors provides a limited calibration range and does not eliminate data distortion across a wide range.
The latter approach relies on a mechanical potentiometer to generate loads across a range. 
A mechanical potentiometer varies its resistance as a result of angular movement.
Although some COTS potentiometers offer the ability to withstand large voltage and current values, they cost more than \$100~\cite{pot}.
More importantly, as mechanical potentiometers are analog devices, the actual changes in resistance cannot be controlled in a discrete manner.  
Therefore, it is difficult to match the readings of a potentiometer versus a baseline.
Furthermore, mechanical potentiometers are prone to performance changes and reliability concerns over time, as they are sensitive to shocks and vibrations.

In addition to the aforementioned approaches, an another solution is to use a commercial DC power analyzer~\cite{lim2013flocklab,potsch2017efficient}, such as Agilent N6705X~\cite{agilent}.  
This device has the capability of acting as a target that generates variable current draws and voltage outputs.
It has a built-in meter and a logger to capture measurement results.  
N6705X can produce accurate and stable current and voltage outputs.  
However, the cost is more than \$14,000 for a complete solution.

In this paper, we present a power calibration tool, named ProCal, which is a scalable tool for calibrating analog-to-digital interfaces.
ProCal is a low-cost and programmable board that provides a wide range of discrete voltage and current outputs for calibration.  
The basic idea of ProCal is to combine a digital potentiometer with a controllable resistor network.
In particular, the digital potentiometer is used for generating small changes in resistance, and the resistor network is controlled by digital switches to enable large resistance jumps.
This is a low-cost approach to overcome the limited range of digital potentiometers currently available.
Therefore, one of the key advantages of ProCal is its large dynamic output range for both current and voltage.
In addition, the entire ProCal tool only costs about \$100, including manufacturing. 
This is less than 2\% of the cost compared to current commercial solutions.

The operations of the digital potentiometer and the resistor network are controlled by software.
This software programs ProCal's output transitions and records the output settling instances.  
If a new current or voltage value is set at a time $t$, the software records $t+ \frac{T}{2}$ as the time stamp of output settling time, where $T$ is the reconfiguration period.
Recording the settling time provides the ability to ensure the readings of the target IoT device and measurement equipment, such as the digital multimeter (DMM) are used as the ground truth and correlate with a stable value, without requiring a precise time synchronization.
This feature notably simplifies calibration because various IoT devices demonstrate different setup times and sampling rates, depending on factors such as ADC speed, processor speed, and available memory.
For example, the sampling rate of an ADC depends on its serial interface, device driver, and programming techniques used.
Therefore, when ProCal simultaneously triggers a DMM and an IoT device to sample their analog inputs, we do not need to worry about the exact start times and sampling rates when matching the pairwise values.

One of the salient features of ProCal is providing output scalability to support a wide range of calibration requirements.
Precisely, the range of the digital output must match the scale of analog input.
For example, assume that an ADC can measure between 0 to 5V, however, the light sensor connected to ADC has sensing range between 0 to 3V, thus creating a void from 4 to 5V.
Therefore, the digital output range needs to map between analog input from 0 to 3V to maximize the resolution of calibration.
To show the scalability of ProCal, we incorporate the Fibonacci sequence to formulate a mathematical model of scalability.
Therefore, ProCal enables the user to adjust resolution depending on the application at hand.

We have implemented a prototype of ProCal and conducted an extensive set of experimental evaluations.
Our results show that ProCal provides satisfactory measurement fidelity under a wide range of operations.  
In particular, ProCal produces a dynamic current range 0.476mA to 1A, and dynamic voltage range 0.06mV to 5V. 
The minimum resolutions of current and voltage are 1.82$\mu$A and 0.01$\mu$V, respectively.
We also present case studies where ProCal is used for calibrating high-resolution ADCs.  
Our calibration shows improvement across different devices.
For example, for ATMega2560 the ADC error reduces from 0.2\% to 0.01\%, and for INA219 the ADC error reduces from 0.42\% to 0.02\%.

The rest of the paper is organized as follow. 
Section~\ref{design} presents the design challenges of scalable calibration, as well as the design, mathematical modeling, and implementation of ProCal.
Evaluation of ProCal through various case studies is given in Section~\ref{eval}.
Section~\ref{relate} discusses the related work. 
We conclude the paper in Section~\ref{conclusion}.

\section{Design Challenges, Considerations and Implementation} 
\label{design}

The primary objective of designing a calibration tool is to calibrate the analog inputs of IoT devices as well as those devices that perform data conversion between digital and analog, such as analog-to-digital converter or digital-to-analog converter.
However, designing a scalable calibration tool poses the following challenges:
\begin{itemize}
    \item \textbf{Time synchronization.}  
    The power calibration tool should provide a time synchronization mechanism between the target device and a high precision measurement equipment, such as DMM or oscilloscope.
    Otherwise, target and DMM may be calibrating values belonging to different voltage or current settings.
    With measurement equipment sampling rates of 1MHz, these event samples must be time stamped within 1$\mu$s accuracy or better.
    
    \item \textbf{High accuracy and dynamic range.}  
    The power calibration tool must provide highly accurate current and voltage samples over a dynamic range that spans five orders of magnitude in current draw or voltage drop.  
    The changes in the current or voltage events must be stable, otherwise pairwise correlation of values would be meaningless.
    
    \item \textbf{Volume calibration steps.}  
    One of the challenges for data converter calibration is its non-linear distortion.
    A well-characterized voltage profile can accomplish the correction of data distortion.
    The power calibration tool should provide high resolution current and voltage change steps to give full coverage for data converter characterization. 
    
    \item \textbf{Portability and low cost.}  
    Compared to high-cost commercial power analyzers, the tool should be easy to integrate, both electronically and mechanically.  
    It should also be less expensive and easy to use, compared to commercial products. 
\end{itemize}

In the rest of this section, we present our design and formulate mathematical models to prove the operating range and resolution of ProCal.
We then present the hardware and software implementation of ProCal.

\subsection{Design Considerations and Solutions}

\subsubsection{Potentiometer Selection.}

\begin{table}[t]
\caption{Comparison of mechanical and digital potentiometers.  Digital potentiometers provide resistance resolution that is programmable and repeatable.}\label{tab:a}
\scriptsize
\centering
{\def\arraystretch{1.1}
\begin{tabular} { |c|c|c|c|c| } 
 \hline
 \textbf{Potentiometer Type }& \textbf{Resolution} & \textbf{Programmable} & \textbf{Accuracy} & \textbf{Operating Range}\\
 \hline
 \hline
Mechanical  & Infinite  &  None & Low & Wide Range \\\hline
Digital  & Limited  & SPI/I2C & High & Limited Range\\
 \hline
\end{tabular}}
\end{table}

The initial objective for calibration is to generate variable loads to provide current and voltage variations. 
Both digital and mechanical potentiometers can satisfy this objective.
Table~\ref{tab:a} compares these two types based on resolution, programmability, accuracy, and operating range.

Theoretically, the resolution of mechanical potentiometers is infinite.  
In practice, however, the effectiveness of the resolution is determined by the skill level of the potentiometer adjuster.
Besides, mechanical potentiometers use a wiper to contact a resistive element that moves along its length to vary resistance. 
Because of the mechanical contacts involved, the resistive could vary by vibration, shock, and pressure.

On the other hand, digital potentiometers consist of CMOS transmission gates.
Although digital potentiometers cannot offer an infinite resolution, they can provide a resolution that appears to be continuous across the supported range. 
More importantly, resolution of digital potentiometers is fully specified, repeatable, and predictable.
Due to these benefits, as well as their tolerance for vibration, shock, and pressure~\cite{creech2015digital}, our design employs a digital potentiometer.

\subsubsection{Dynamic Output Range.}\label{rangeprove}

The standard analog-to-digital conversion has analog input $V_{in}(t)$ and digital output $D_{out}(t)$, where $t$ represents time.
The purpose of voltage calibration is to find a function $f(d)$ to be implemented in the correction module, where $d$ is digital output.
Function $f(d)$ is to minimize the error between the corrected output $v_c=f(D_{out}(t))$ and ground truth measurement $V_{in}(t)$ at time $t$. 
Similarly, for current calibration, a function $I_{in}(t)$ is derived.

Equation~\ref{eqn:lsb} defines the minimum resolution of ADC per digital bit (LSB), as follows,
  \begin{equation}
  \label{eqn:lsb}
    LSB = \Delta = \frac{V_{FS}}{2^n} 
  \end{equation}
where $V_{FS}$ is the full-scale range of ADC determined by reference voltage and $n$ is the maximum number of bits the input value can be translated to.  
Equation~\ref{eqn:dout} defines the final digital output value of the ADC,

\begin{equation}
\label{eqn:dout}
   D_{out}(t)=\left.\Bigl\lfloor\frac{V_{in}(t)}{\Delta}\Bigr\rfloor\right|_{t=n\times T_s}=\Bigl\lfloor2^n\times\frac{V_{in}(n \times T_s)}{V_{FS}}\Bigr\rfloor
\end{equation}
where $T_{s}$ is sampling period. 
Based on this equation, normalization and truncation of analog inputs are involved in the ADC quantization process~\cite{bennett1948spectra}. 
The maximum instantaneous value of distortion is $\frac{LSB}{2}$, and the total range of variation is from $-\frac{LSB}{2}$ to $+\frac{LSB}{2}$.
In addition to quantization errors, there are two static specifications that define the accuracy of analog-to-digital conversion: \textit{differential non-linearity} (DNL), and \textit{integral non-linearity} (INL).
DNL error is defined as the difference between an actual step width and the ideal value of 1 LSB.
INL error is described as the deviation, in LSB or percentage of full-scale range, of an actual transfer function from ideal straight conversion line.
More importantly, the INL error magnitude directly depends on the correlation position chosen for ideal straight line. 
Since $f(d)$ is non-linear, it is crucial to characterize ADC's full scale instead of only a few positions.  

Typical energy estimation for an IoT device requires the measurement of both current and voltage.
Current range is 0mA to 1A and voltage range is 0mV to 5V (i.e., CMOS operation range).
A calibration tool should generate both variable currents and voltages.
However, the critical challenge is to create various currents in a wide dynamic range.
Specifically, since the digital potentiometer, wiper, and potentiometer connection are limited to the bounds of the power-supply rail, the potentiometer can only carry limited current and voltage.
Most of the existing digital potentiometers support current in the range 0.6 to 3mA, and only a few products that accommodate up to 20mA.
In any case, the potentiometer operating condition must be within the range of CMOS operation. 
This limitation is against the desired current range.
To overcome this challenge, we connect a resistor network in parallel with the digital potentiometer to increase the total current flow. 
We first express the current output of digital potentiometer as an equation.  Then, this equation is incorporated into a mathematical model, which models the scalability of ProCal.
The current output range for a n-bit digital potentiometer can be expressed as follows
\begin{equation}
  \label{eqn:pot_base}
    \left.\frac{V_{in}(t)}{R(2^n)+R_b}\leq I_{pot}(t) \leq \frac{V_{in}(t)}{R(0)+R_b}\right.
\end{equation}
where $t$ is time, $I_{pot}(t)$ is current output of digital potentiometer, $R(0)$ is resistance when digital potentiometer is programmed to 0,
$V_{in}(t)$ is the input voltage, $R_b$ is the resistor connected to digital potentiometer in serial, and $R(2^n)$ is digital potentiometer's maximum resistance value when programmed to $2^n$.
Through connecting $R_b$ with the digital potentiometer's input rail in serial, we protect the digital potentiometer from overcurrent. 
The maximum current the digital potentiometer can support is expressed  as follows
\begin{equation}
  \label{eqn:pot_base_max}
    I_{potmax}(t)=\frac{V_{in}(t)}{R(0)+R_b}=\frac{V_{in}(t)}{R_w+R_b} 
\end{equation}
where $I_{potmax}$ is maximum current digital potentiometer can accommodate, $R_w$ is the constant wiper resistance when digital potentiometer is programmed to 0.

Base on Kirchoff's current law, the current flow into a device equals the current flow out of the device. 
Since digital potentiometer can only accommodate up to $I_{potmax}(t)$ continuous current, we add a resistor network in parallel with the digital potentiometer to increase the total current flow in order to support larger current range.
The total current digital potentiometer and parallel resistor network can support is expressed as follows,
\begin{equation}
  \label{eqn:current_sum}
    \left. I_{ProCal}(t) = I_{potmax}(t) + \sum_{j=1}^n I_j(t) \;\;\;\; \forall n \in Z^+ \right.
\end{equation}
where $I_{ProCal}(t)$ is the total current ProCal can support, $\sum_{j=1}^n I_j(t)$ is the sum of the current that the resistor network can support, and $ I_j(t)$ is the current that a resistor $R_j$ belonging to the resistor network can support.
For any positive integer $n$, we can expand the current output range $I_{ProCal}(t)$ to satisfy any calibration range requirement as long as the given input voltage $V_{in}(t)$ is within the operation range of digital potentiometer and resistor network.
With the capability to expand the output range, the current scalability of ProCal is proven.

Similarly, for voltage scalability, we connect a resistor $R_c$ with the resistor network in serial. 
Based on Ohm's law, ProCal's voltage output range can be expressed as follows,
\begin{equation}
  \label{eqn:voltage_sum}
    \left. V_{ProCal}(t) = I_{ProCal}(t)\times R_c\right.
\end{equation}
where $V_{ProCal}(t)$ is the total voltage ProCal can support.
With the capability to expand the $I_{ProCal}(t)$ range, we proved the voltage scalability of ProCal. 

In addition to scalability, we want the ProCal's current and voltage output to be as close as possible to a linear function. 
Because ADC nonlinearity requires curve-fitting to achieve accuracy, we can improve calibration accuracy by providing a linear output over a wide range.
In order to generate a linear output, $I_j(t)$ should have a linear relationship with $I_{potmax}(t)$.
The linear relationship between $I_j(t)$ and $I_{potmax}(t)$ is expressed as follows
\begin{equation}
  \label{eqn:linear}
   \left. I_j(t) = k\times I_{potmax}(t) \;\;\;\; \forall k \in N \right.
\end{equation}
where $I_{potmax}(t)$ is the maximum current of digital potentiometer, and each $I_j(t)$ current that resistor network can support is $k$ times of $I_{potmax}(t)$.  
To support gradual increase of output current, $I_j(t)$ can be expressed as follows
\begin{equation}
  \label{eqn:fibnacci}
    I_j(t) = I_{j-1}(t)+I_{j-2}(t)
\end{equation}
where every $I_j(t)$ is the sum of previous two terms, also known as the Fibonacci sequence.
Therefore, we use induction to prove the maximum current as shown below
\begin{equation}
  \label{eqn:fibnacci_sum}
   \left. \sum_{j=1}^n I_j(t) = I_{n+2}(t) - 1 \;\;\;\; \forall n\ge2\right.
\end{equation}

With the mathematical model proven, we implement our prototype with $n=3$. 
However, based on Equation~\ref{eqn:current_sum}, scalability of ProCal is not necessarily limited to $n=3$.
We select digital potentiometer AD5200~\cite{AD5200} to generate $I_{pot}(t)$ current. 
The reason behind selecting AD5200 is that it supports up to 20mA, which is larger than most of the digital potentiometers currently available in the market. 
According to Equation~\ref{eqn:pot_base}, we connect a $220\Omega$ resistor with AD5200 in serial as overcurrent protection resistor $R_b$.

Since $I_{potmax}(t)= I_1(t) = 20$, we choose $I_2(t) = 80 $ and $I_3(t) = 100 $ to implement our prototype, where $I_3(t) = I_1(t) + I_2(t)$ satisfies Equation~\ref{eqn:fibnacci},
$I_2(t) = 4 \times I_{potmax}(t)$ and $I_2(t) = 5 \times I_{potmax}(t)$ satisfy the linearity relationship of Equation~\ref{eqn:linear}.
We use ADG1612~\cite{ADG1612} digital switches to control the number of resistors in the resistor network that are connected to the circuit to provide more significant current jumps, compared to what is provided by AD5200.
In particular, initially, the digital potentiometer is used for introducing small changes in the current by programming the resistance value from high to low.
After reaching the maximum current of digital potentiometer, a resistor on the resistor network,  controlled by the digital switch, is enabled to provide additional current.  
Repeating this process results in gradually increasing the current through enabling additional resistors until reaching ProCal's maximum current output.  
Based on Equation~\ref{eqn:voltage_sum}, a similar analysis can be employed for voltage.

\subsubsection{Validation of Minimum Resolution.}
The minimum resolution of our design depends on the digital potentiometer's maximum end-to-end resistance $R_{max}$ and the number of bits supported by digital potentiometer. 
The equation that determines the digitally-programmed output resistance is expressed as follows
\begin{equation}
  \label{eqn:digipot resistor}
   \left. R(x) = \frac{x}{2^n}\times R_{max}+R_w  \;\;\;\;  0\leq x \leq 2^n \right.
\end{equation}
where $n$ is the number of bits supported by digital potentiometer, $x$ is the value programmed into the digital potentiometer, and $R_w$ is the wiper resistance.
Output resistance $R(x)$ is proportional to the digital potentiometer's end-to-end resistance $R_{max}$.

The minimum resolution of current output can be expressed as follows
\begin{equation}
  \label{eqn:minimum res}
   \left.  I_{res}(t) = \frac{(R(x)-R(x-1))}{R(x)\times R(x-1)} \times V_{in}(t) \;\;\;\; 0 < x\leq 2^n \right.
\end{equation}
For example, ProCal uses an 8-bit digital potentiometer, AD5200 which has maximum end-to-end resistance $10k\Omega$. 
Based on Equation~\ref{eqn:digipot resistor}, resistance $R(256)$ is $10.282K\Omega$ and $R(255)$ is $10.242k\Omega$.
Considering ProCal has input voltage 5V, and a $220\Omega$ resistor connected to the digital potentiometer in serial, we can calculate ProCal's minimum current output as $0.476mA$ and the minimum current resolution as $1.82\mu$A, based on Equation~\ref{eqn:minimum res}.
Minimum current resolution and output current range can be further reduced or increased by choosing different specs of the potentiometer or supply voltage.
A similar approach is used to calculate ProCal's minimum voltage resolution and voltage output range.
We validated that ProCal's minimum voltage output is $0.06mV$ and minimum voltage resolution is $0.01\mu$V.

\subsubsection{Time Synchronization.}
The differences in the sampling offset and sampling rates of DMM and target device make correlating the readings a challenging task.
For example, when an output value is configured, ADC might miss that value due to its lower sampling rate compared to DMM.  
If the design does not provide enough duration during two consecutive output changes, ADC and DMM can collect samples belonging to different output values.
Therefore we need to ensure that both DMM and IoT device sample output at least once during a configuration period.
The next challenge is to ensure that when we compare two sampled values, those belong to the same interval when the output was stable.
Therefore, ProCal should also provide mechanisms to synchronize the readings of ADC and DMM.
To overcome these challenges, ProCal stores voltage or current settling time information after applying each configuration that results in current or voltage change.
For example, assume that the minimum sampling rate of ADC and target device is $r_{min}$.
We first ensure that $T > \frac{1}{r_{min}}$, where $T$ is the period of changing the output.
ProCal provides this feature to support the sampling rate of various ADCs.
Second, assuming that a new output value is configured at time $t_i$, to correlate the values collected by DMM and target IoT device, we find in their readings the values that are closest to time $t_i + \frac{T}{2}$.
This synchronization approach ensures that the two values belong to the same output value.

\begin{figure}[!t]
\centering
  \includegraphics[width=0.8\linewidth]{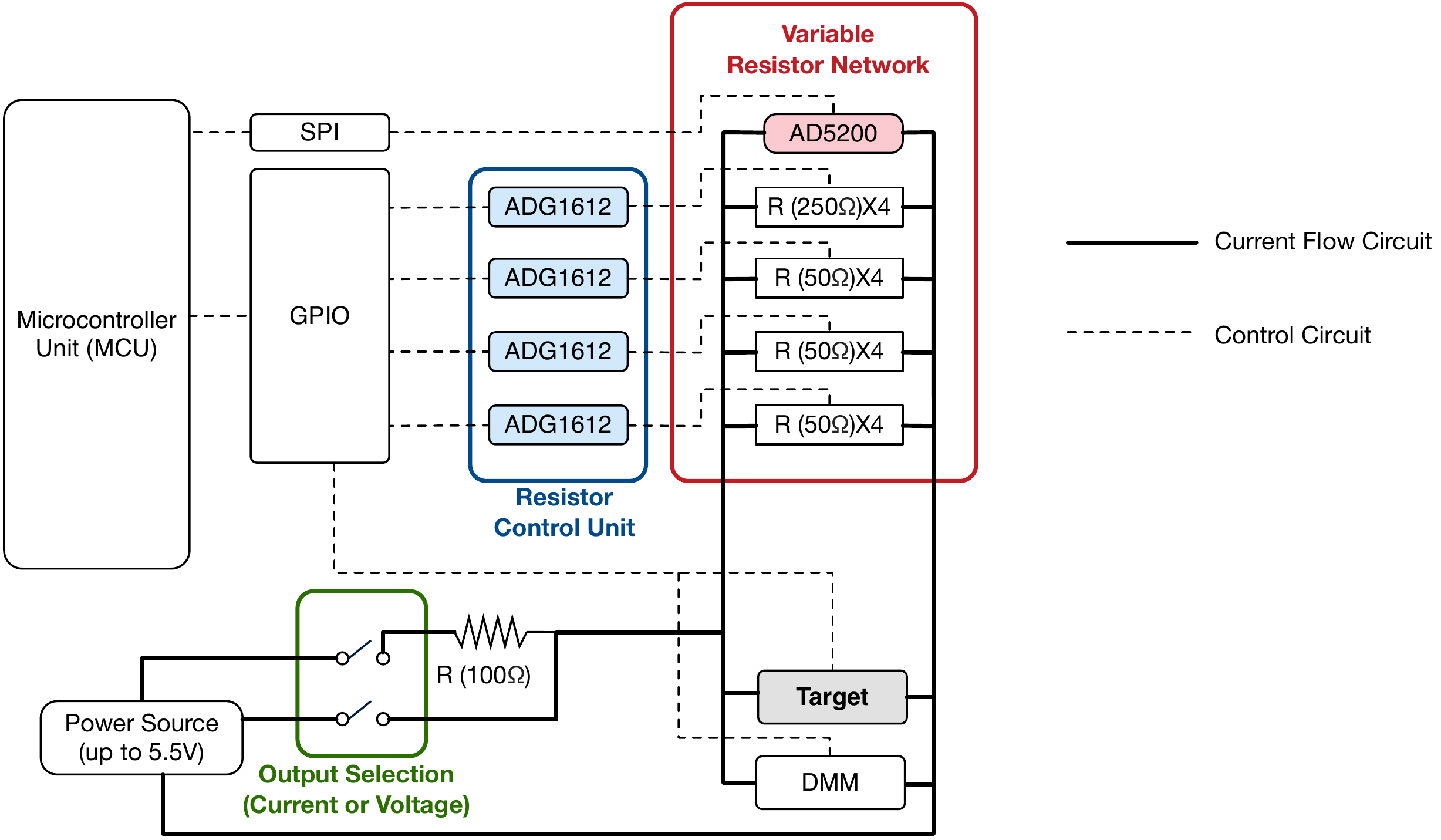}
  \caption{Block diagram of ProCal. MCU provides synchronized trigger signals to the resistor control unit, target ADC, and DMM. AD5200 is a digital potentiometer, and ADG1612 is the digital switch connected to the resistor network.}
  \label{fig:EMCP_overview}
\end{figure}

\subsection{Hardware Implementation}

ProCal's hardware consists of four main units: (i) microcontroller unit (MCU), (ii) resistance control unit (RCU), (iii) variable resistor network, and (iv) output selection jumpers.  
MCU is used to control RCU, synchronize the operation of the target device and DMM, and record settling time of each output value generated by ProCal.   
RCU is responsible for controlling the variable resistor network. 
The output selection jumpers on the board enable the user to select current or voltage output for calibration.

\begin{figure}[!t]
\centering
  \includegraphics[width=0.67\linewidth]{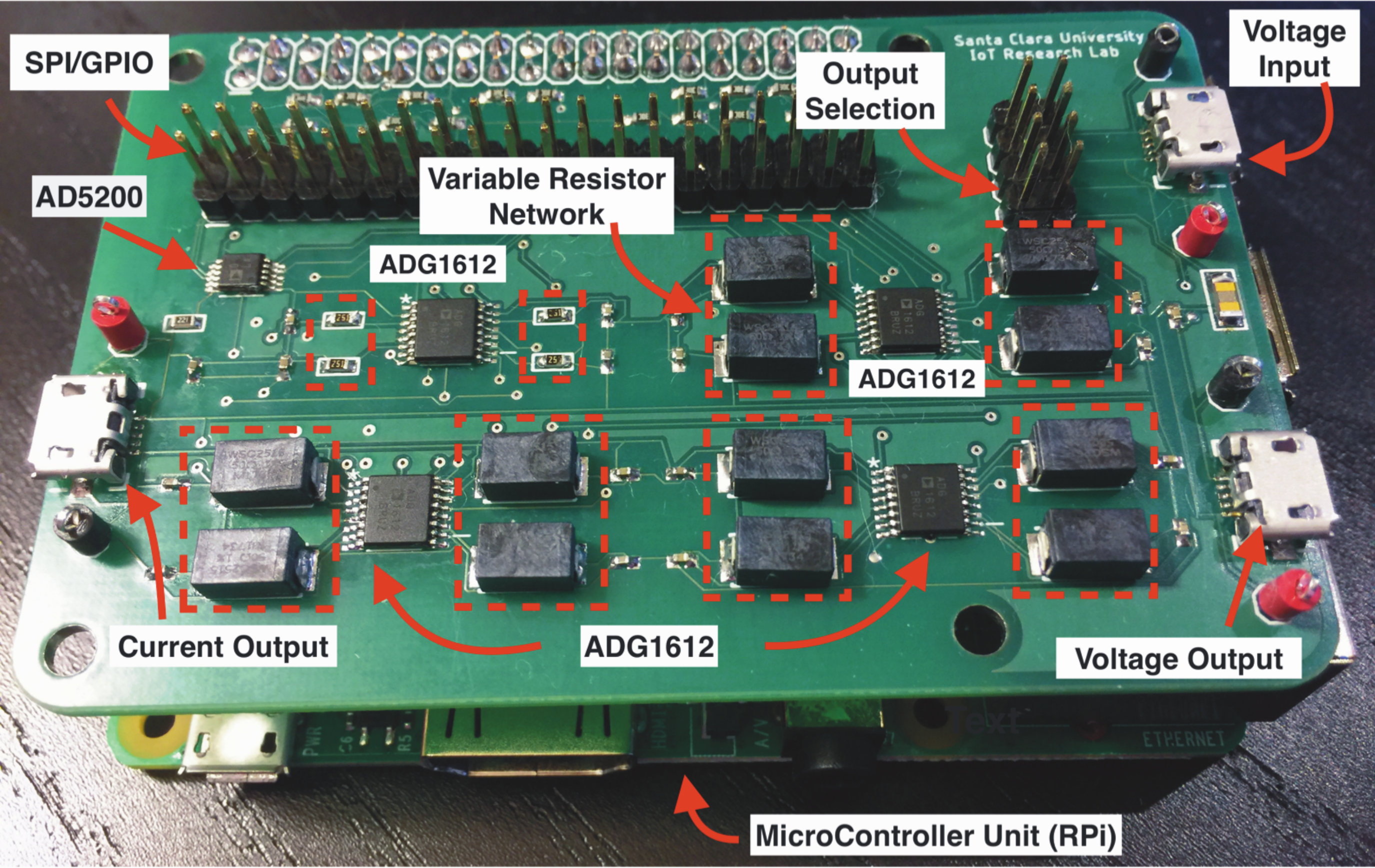}
  \caption{ProCal prototype. ProCal is installed on top of a RPi. The RPi communicates with ProCal through SPI and GPIO interfaces. The dashed lines show the resistor network.}
  \label{fig:EMCP_prototype}
\end{figure}

Figure~\ref{fig:EMCP_overview} presents the block diagram of ProCal. 
Figure~\ref{fig:EMCP_prototype} shows a ProCal board.  
We implement MCU function by developing a Python configuration code running on a Raspberry Pi 3 board.  
MCU connects to RCU through a SPI~\cite{semiconductor2014official} bus and GPIO pins. 
MCU controls the RCU to perform the desired changes.  
MCU also provides a trigger signal through GPIO to trigger target device and DMM to start and stop sampling simultaneously.

RCU contains one AD5200~\cite{AD5200} digital potentiometer and four ADG1612~\cite{ADG1612} digital switches.
AD5200 offers 256 different resistance values that can be digitally programmed by the configuration code through SPI. 
These different resistances provide full calibration coverage, from a low sleeping current to a high active current of IoT device, as we proved in Section~\ref{rangeprove}. 
We connect a 220$\Omega$ resistor with AD5200 in serial as the overcurrent protection. 
The salient features of AD5200 are its short setup time and 50MHz SPI speed. 
These features enable ProCal to provide stable output current and voltage within 1$\mu$s period.

The variable resistor network is a circuit connecting 16 resistors and a digital potentiometer, in parallel.
ADG1612 contains four terminals, where each terminal connects to a resistor. 
MCU sends logic signals to ADG1612 terminals through GPIO.
When the terminal receives a logic 1 signal, the resistor controlled by the terminal is connected to the variable resistor network, which increases the total current flow. 
If the terminal receives a logic 0 signal, the resistor is disconnected from the variable resistor network, which reduces the total current flow.
There are four ADG1612s used in the RCU: one is connected to four 250$\Omega$ high-precision (1\%) resistors and the other three are connected to four 50$\Omega$ high-precision (1\%) resistors.
These high-precision resistors connected to ADG1612 are part of the variable resistor network, which can provide 20mA and 100mA changes in current. It is worth noting that the effect of ADG1612 on current draw is minimal as it only shows a 1$\Omega$ resistance when operating.

The output selection jumpers are used to switch ProCal's calibration output between voltage and current.
When voltage output is selected, output selection jumper enables a high power 100$\Omega$ resistor to connect in serial with the resistor network.   
It provides the same dynamic range for the voltage change between 0 to 5V as we proved in Section~\ref{rangeprove}.  
ProCal generates trigger signals through GPIO to start and stop data collection of target and DMM simultaneously. 
To improve stability and remove the alternating current caused by ripple voltage, we add low pass filters to the resistor network.

\begin{algorithm}[!t]
\footnotesize

\SetKwFunction{Fmain}{main}
\SetKwProg{Fn}{function}{}{}

\BlankLine

\Fn{\Fmain{$T$, $I_{max}(t)$, $V_{min}(t)$, $V_{in}(t)$, $n$}}
{
    initialize GPIO and disconnect all resistors from the resistor network\;
    setup SPI driver and bus speed to 50MHz\;
    send start trigger to target through GPIO\;
    \textit{breakflag} = 0\;
    \BlankLine
    \For{$i \leftarrow$ 0 \KwTo number of 50$\Omega$ resistors}{
        \uIf{\textit{breakflag} == 1}        {break\;}
        \uIf{$i$ != 0}
            {connect 50$\Omega$ terminal[$i$] to the resistor network\;}
        \For{$j \leftarrow$ 0 \KwTo number of 250$\Omega$ resistors}{
           \uIf{\textit{breakflag} == 1}
                { break\;}
            \uIf{ $i$ != 0 and $j$ == 0 }
                  {disconnect all 250$\Omega$ from the resistor network\;}
            \uElseIf{ $j$ != 0}
                { connect 250$\Omega$ terminal[$j$] to the resistor network\;}
            \For{ $x \leftarrow$ $2^n$ \KwTo 0 }
                {
                 $R(x)$ = $\frac{x}{2^n}\times R_{max}+R_w+R_b$\;
                 $I(t)$ = $\frac{V_{in}(t)}{R(x)}+i\times100+j\times20$\;
                 $V(t) = R_c\times I(t)$\;
                 \uIf{$I(t) \geq I_{max}(t)$ }{
                    \textit{breakflag} = 1\;
                    break\;}
                \uIf{$V(t) \leq V_{min}(t)$ }{
                    \textit{breakflag} = 1\;
                    break\;}
                 program AD5200 with value $x$\;
                 delay for $\frac{T}{2}$\;
                 write time stamp $\frac{T}{2}$, $I(t)$ and $V(t)$ into file\;
                 delay for $\frac{T}{2}$\;

                }
     
           }
   
        }
    send stop trigger to target and DMM through GPIO\;
    \KwRet\;
}

\caption{Configuration Software of ProCal} 
\label{alg:emcp}
\end{algorithm}

\subsection{Software Implementation}

ProCal's software runs on a flavor of Linux operating system called Raspbian and uses BCM and SPI library. 
The operations of the digital potentiometer and the resistor network are controlled by the software.
This is the basis for setting up SPI and GPIO drivers, accurate time stamping of resistance change events, and sending synchronization signals. 
The software uses a collection of Python scripts that are used by MCU to trigger scheduled actions to start and stop a calibration process.
The software programs the AD5200 and ADG1612 to provide output transitions, and records the settling time between two consecutive output changes.

As shown in Algorithm~\ref{alg:emcp}, the controller software first initializes all GPIO ports, sets up the SPI driver, and adjusts the SPI clock speed to 50MHz. 
Users need to provide: (i) a desired time interval between configurations $T$, (ii) the maximum current $I_{max}(t)$ or minimum voltage $V_{min}(t)$ to calibrate, (iii) supply voltage to ProCal $V_{in}(t)$, and (iv) number of bits the digital potentiometer can support.  
ProCal can support output frequency (i.e., changes of current/voltage values) up to 50MHz.
After MCU sends the trigger signals to target device and DMM through GPIO, it starts to change resistance until the desired maximum current or minimum voltage is reached. 
The software time stamps every current/voltage change events and saves them into a log file.
This log is used for time synchronization between the target device and DMM measurements.

\section{Performance Evaluation and Case Studies} \label{eval}
In this section, we evaluate ProCal's dynamic range, resolution, and stability.
We show the effectiveness of ProCal to address the most challenging requirements of calibrating IoT devices. 
In particular, we present case studies of using ProCal to calibrate voltage and current with commercial ADCs.

\subsection{Evaluation of Dynamic Range}
Dynamic range and resolution are important performance metrics for a calibration platform. 
ProCal's output range and resolution can be customized to within the target ADC's range of interest. 
To measure dynamic range, resolution, and static accuracy, a high accuracy DMM, Keithley DMM7510~\cite{k7510}, is used to validate the results.  
DMM7510 provides picoampere-level sensitivity and sampling rate 1Msps, which can precisely validate the performance of ProCal.  
When validating dynamic range, we connected DMM to ProCal's output to measure ground-truth current and voltage.  
For this experiment, output current changes every 5ms from 0.4mA to 900mA and voltage changes every 5ms from 5V to 0.06mV.

\begin{figure}[!t]
\centering
  \includegraphics[width=0.8\linewidth]{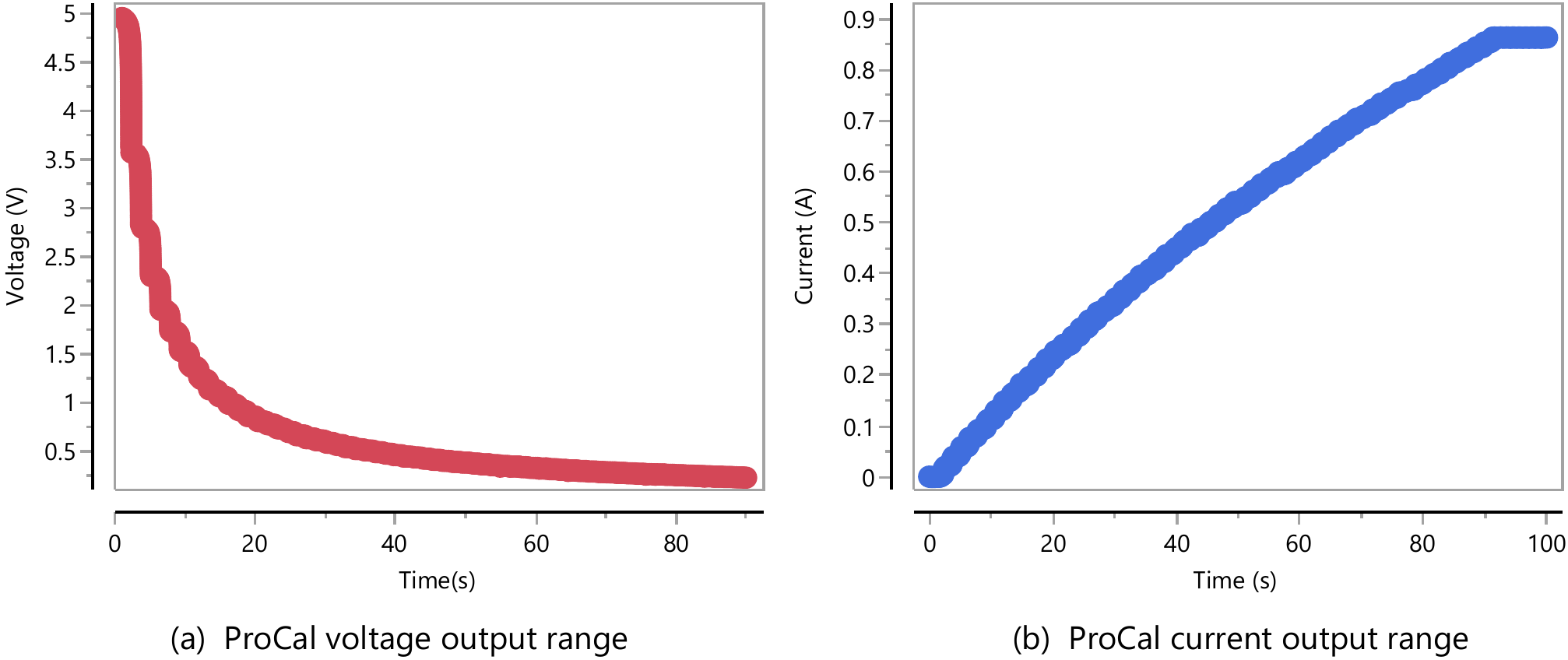}
  \caption{Experimental results of ProCal prototype. (a) ProCal voltage output. Our prototype supports voltage output from 0.06 mV to 5 V. (b) ProCal current output. Our prototype supports current output from 0.476 mA to 900 mA. These current and voltage values address the requirements of calibrating various types of IoT devices.}
  \label{fig:EMCP_range}
 \end{figure}

Figure~\ref{fig:EMCP_range} shows the experimental results pertaining to ProCal's dynamic range validation. 
It can be seen that ProCal output current spans from 0.5mA to 900mA and voltage spans from 5V to 0.06mV.  
The results in this section validate the wide dynamic range and high resolutions.

\subsection{Case Studies}
In this section, we present case studies to demonstrate the benefits of calibration using ProCal. 
In these case studies, we use ProCal to perform real-world calibration of the most popular COTS ADCs, namely ATMega2560~\cite{Arduino}, MCP3208~\cite{MCP}, and INA219~\cite{ina219}. 
Our ground truth measurement is obtained by a high-accuracy DMM, Keithley DMM7510.
In particular, we evaluate the performance of current and voltage calibration. 
To this end, we use ProCal to generate a dynamic range of voltage and current values.  
We use the target ADC and DMM to measure the output of ProCal simultaneously.
After completing the measurements, we use the saved time stamp log to correlate the measurements between ADC and DMM. 
For each experiment, we generate a calibration function $f(d)$ or a calibration table to correlate ADC and DMM measurements.
Note that as the log file reflects the stability time stamp of each configuration, we can safely compare two closest entries of the traces collected by ADC and DMM without requiring precise time synchronization of the two devices.

\begin{figure}[!t]
\centering
     \includegraphics[width=0.8\linewidth]{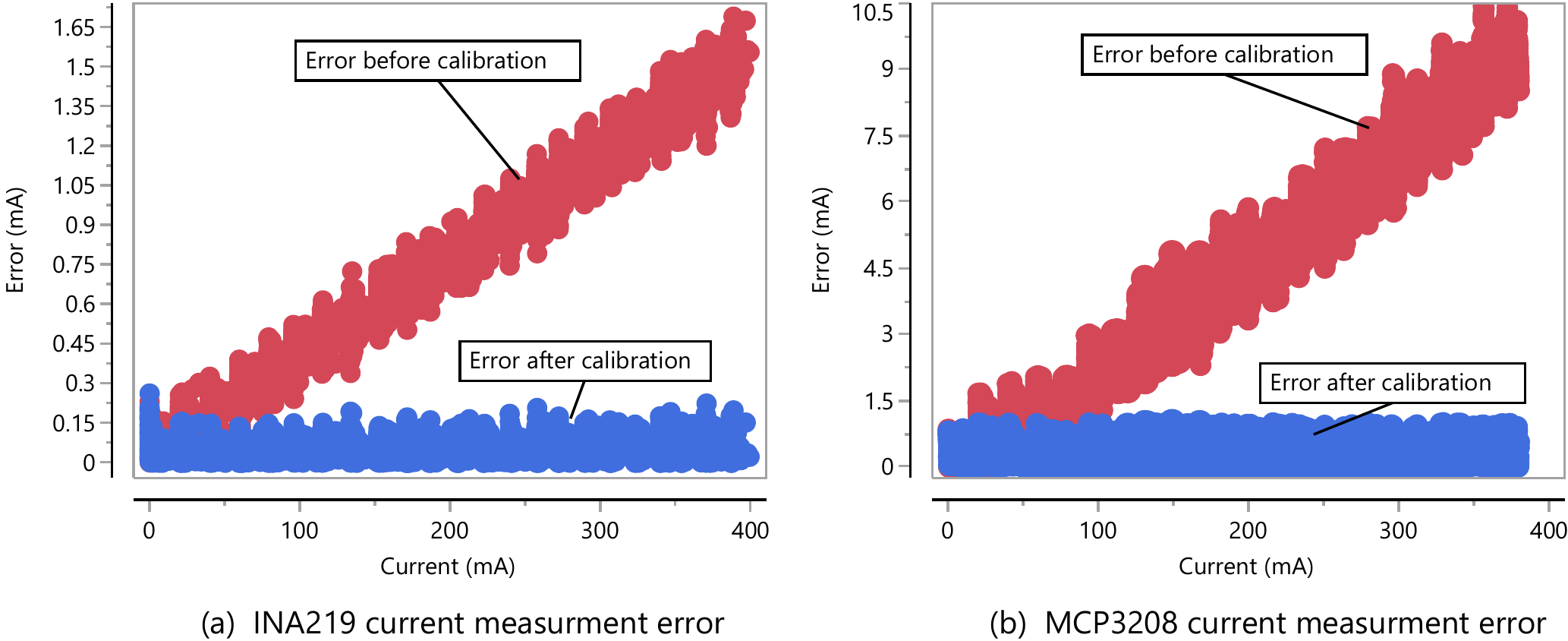}
  \caption{Current calibrations using ProCal. Experimental results for (a) INA219, and (b) MCP3208 calibrations. These results show current measurement errors before and after calibration. Errors are significantly reduced after calibration. }
  \label{fig:ina_current}
 \end{figure}

Figure~\ref{fig:ina_current} shows the current measurement error of INA219 and MCP3208 conducted in a normal indoor temperature  25\textdegree{C}.
As it can be observed, the error of both ADCs increases linearly versus current. 
Therefore, we find the best polynomial fitting curve to calibrate the errors.

\begin{figure}[!t]
\centering
  \includegraphics[width=0.8\linewidth]{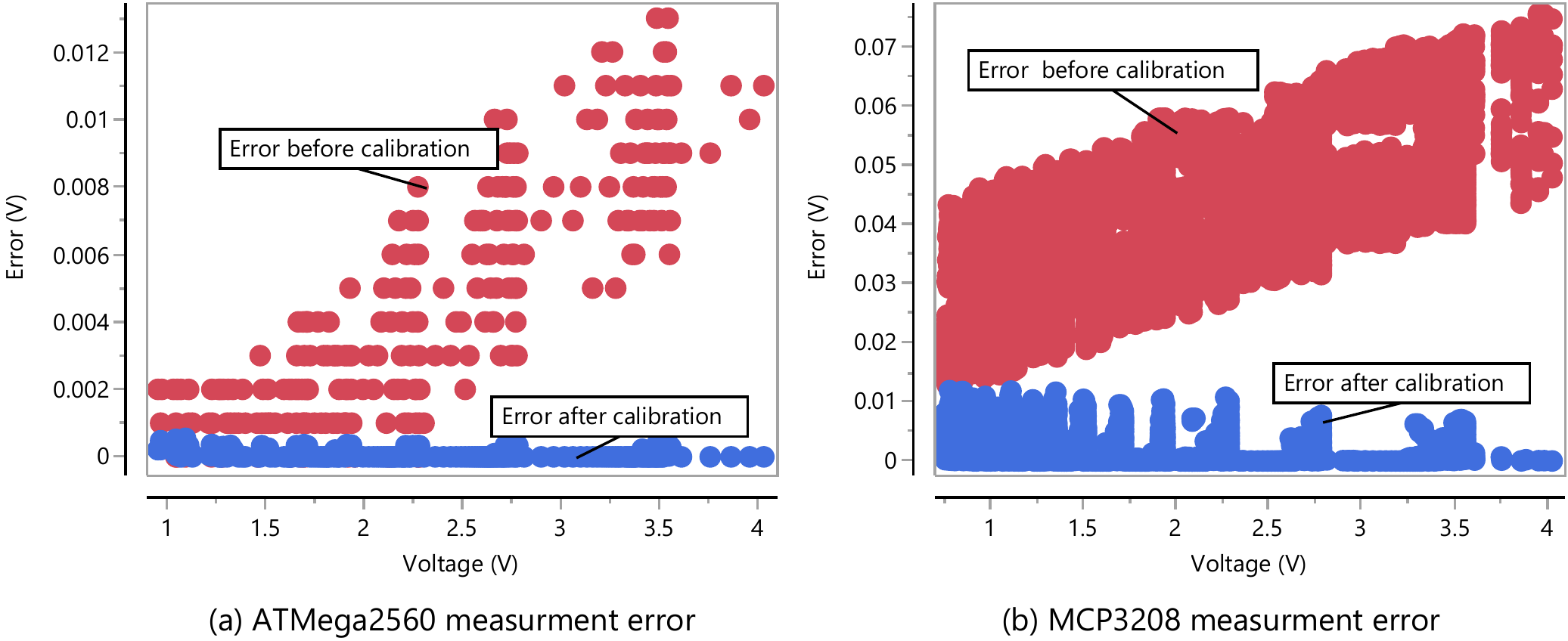}
  \caption{Voltage calibrations using ProCal. Experimental results for (a) ATMega2560, and (b) MCP3208 calibrations. These results show voltage measurement error before and after calibration.}
  \label{fig:mcp_voltage}
 \end{figure}

Figure~\ref{fig:mcp_voltage} shows the voltage measurement error of ATMega2560 and MCP3208.
The error of both ADCs increases non-linearly versus voltage. 
For these cases, we used a calibration table to find the best fit of the ADC measurement.

\begin{table}[!t]

\caption{Comparison of measurement errors before and after calibration}
\label{tab:b}
\scriptsize

\centering
{\def\arraystretch{1.3}
\begin{tabular} { |c|c|c|c| }
 \hline

 \textbf{ADC} & \textbf{Measurement Type} & \textbf{\%Error (before) } & \textbf{\%Error (after)}\\
 \hline
 \hline

 INA219  & Current  &  0.42\% & 0.02\% \\
  \hline
 MCP3208 & Current  &  2.58\% & 0.01\% \\
 \hline
 MCP3208  & Voltage  &  5.29\% & 0.01\% \\
  \hline
 ATMega2560  & Voltage  &  0.2\% & 0.01\% \\
 \hline

\end{tabular}}
\end{table}

Table~\ref{tab:b} presents the calibration results, and confirms the significant effect of ProCal calibration on measurement accuracy.
These results demonstrate that by providing an extensive calibration range, the errors are reduced to nearly 0.01\%.
For example, in the MCP voltage case, the ADC error reduces from 5.29\% to 0.01\%.

\section{Related Work}
\label{relate}
Data conversion accuracy is one of the main requirements of IoT applications.
However, there are various factors affecting the precision of data conversion.  
For example, device characteristic, process technology, resistance of the circuit, temperature variations, and finite gain, are among the parameters that affect accuracy.
Standard methods to minimize the effects of these parameters are \textit{trimming} and \textit{calibration}.
Trimming is performed during the production phase.  
After measuring and correcting the parameters of the device at a controlled condition, the trimmed values are programmed into the device. 
Trimming values, however, cannot be changed after manufacturing.  
Therefore, the subsequent drift due to device aging, temperature, or system-level noise, cannot be corrected.

Calibration can be performed multiple times after the device is fabricated. 
Therefore, it can be used to compensate for changes that occur over time and those influenced by environmental conditions.
There are two types of calibrations: \textit{self-calibration}, and \textit{external calibrations}.
Self-calibration performs the measurement-correction process within the device itself and does not require an external device.
The calibration process that interrupts the ordinary data conversion process and is performed during device power up is referred to as foreground calibration or power up calibration~\cite{taft20041,lee1984self,261994,cao2016power}.
Although this technique results in significant error reduction, it cannot cope with temperature drift and other sources of inaccuracy such as sensor-to-ADC path loss.
Alternatively, data converter can be calibrated while it is performing the regular conversion, and the result of conversion would not be affected by the calibration process.
This is referred to as background calibration or runtime calibration~\cite{tsang2008background,moon1997background}.
Background calibration encompasses temperature effect, which provides a higher absolute accuracy than foreground calibration.

However, self-calibration is only as accurate as the onboard reference voltage.  
It can drift slightly over time, and thereby, external calibration is still essential.
For example, although the 12-bit SAR ADC of STM32F205 includes a background calibration feature, Hartung et al.~\cite{hartung2016distributed} used two resistors to perform external calibration, and they showed that voltage and current errors are reduced from 2.36\% and 2.21\% to 0.87\%  and 1.56\%, respectively.
Similarly, Zhou et al.~\cite{zhou2013nemo} relies on external calibration, although TI MSP430F2618 has a background calibration capability.
The common approaches of external calibration include using: (i) fixed resistors~\cite{milenkovic2005environment,jiang2007micro}, (ii) mechanical potentiometer~\cite{zhou2013nemo,haratcherev2008powerbench}, and (iii) commercial power analyzer~\cite{lim2013flocklab,potsch2017efficient}.
However, as we mentioned in Section~\ref{intro}, these calibration mechanisms pose several limitations.
Resistors and potentiometers cannot provide quality calibrations due to limited range or unreliable resistance. 
Furthermore, although commercial power analyzers can provide quality results, they are costly.
The proposed ProCal tool provides a cost-efficient solution without imposing any of the limitations of the aforementioned solutions.

\section{Conclusion} 
\label{conclusion}
Calibration is an important step towards building reliable IoT systems.
For example, accurate sensor reading requires ADC calibration, and power monitoring chips must be calibrated before being used for measuring the energy consumption of IoT devices.
In this paper, we presented a power calibration tool, called ProCal, which is a programmable and scalable tool.
ProCal is a low-cost programmable platform that provides dynamic voltage and current output for calibration. 
The basic idea is to use a digital potentiometer connected to a parallel resistor network controlled through digital switches.
The resistance and output frequency of ProCal is controlled by a software communicating with the board through SPI and GPIO interfaces. 
We incorporated the concept of Fibonacci sequence into our mathematical model to prove that ProCal can be extended to other and wider ranges.
Our extensive experimental case studies demonstrated that ProCal can reduce measurement errors significantly.
In addition to using ProCal for calibration, it can be employed for power emulation. 
Specifically, instead of using an IoT board to generate sudden variations in current, ProCal can be used to emulate the operation of an IoT device under various conditions.
ProCal can also be used, for example, to test the resilience of a power harvesting system under various types of loads.

\bibliography{bibliography}

\begin{thebibliography}{10}
\providecommand{\url}[1]{\texttt{#1}}
\providecommand{\urlprefix}{URL }

\bibitem{bennett1948spectra}
Bennett, W.R.: Spectra of quantized signals. Bell Labs Technical Journal
  27(3),  446--472 (1948)

\bibitem{bychkovskiy2003collaborative}
Bychkovskiy, V., Megerian, S., Estrin, D., Potkonjak, M.: A collaborative
  approach to in-place sensor calibration. In: Information Processing in Sensor
  Networks. pp. 301--316. Springer (2003)

\bibitem{cao2016power}
Cao, J., Meng, X., Temes, G.C., Yu, W.: Power-on digital calibration method for
  delta-sigma adcs. In: IEEE International Symposium on Circuits and Systems
  (ISCAS). pp. 2002--2005. IEEE (2016)

\bibitem{creech2015digital}
Creech, J., Rice, D.: \relax{Digital potentiometers vs. mechanical
  potentiometers: Important design considerations to maximize system
  performance}. Analog Devices, MA, USA, Technical article  (2015)

\bibitem{dezfouli2018empiot}
Dezfouli, B., Amirtharaj, I., Li, C.C.: Empiot: An energy measurement platform
  for wireless iot devices. arXiv preprint arXiv:1804.04794  (2018)

\bibitem{dezfouli2017rewimo}
Dezfouli, B., Radi, M., Chipara, O.: Rewimo: A real-time and reliable low-power
  wireless mobile network. ACM Transactions on Sensor Networks (TOSN)  13(3),
  ~17 (2017)

\bibitem{haratcherev2008powerbench}
Haratcherev, I., Halkes, G., Parker, T., Visser, O., Langendoen, K.:
  Powerbench: A scalable testbed infrastructure for benchmarking power
  consumption. In: Int. Workshop on Sensor Network Engineering (IWSNE). pp.
  37--44 (2008)

\bibitem{hartung2016distributed}
Hartung, R., Kulau, U., Wolf, L.: Distributed energy measurement in wsns for
  outdoor applications. In: 13th Annual IEEE International Conference on
  Sensing, Communication, and Networking (SECON). pp. 1--9. IEEE (2016)

\bibitem{jiang2007micro}
Jiang, X., Dutta, P., Culler, D., Stoica, I.: Micro power meter for energy
  monitoring of wireless sensor networks at scale. In: Proceedings of the 6th
  international conference on Information processing in sensor networks. pp.
  186--195. ACM (2007)

\bibitem{261994}
Karanicolas, A.N., Lee, H.S., Barcrania, K.L.: A 15-b 1-msample/s digitally
  self-calibrated pipeline adc. IEEE Journal of Solid-State Circuits  28(12),
  1207--1215 (12 1993)

\bibitem{lee1984self}
Lee, H.S., Hodges, D.A., Gray, P.R.: A self-calibrating 15 bit cmos a/d
  converter. IEEE Journal of Solid-State Circuits  19(6),  813--819 (1984)

\bibitem{lim2013flocklab}
Lim, R., Ferrari, F., Zimmerling, M., Walser, C., Sommer, P., Beutel, J.:
  Flocklab: A testbed for distributed, synchronized tracing and profiling of
  wireless embedded systems. In: Proceedings of the 12th international
  conference on Information processing in sensor networks. pp. 153--166. ACM
  (2013)

\bibitem{milenkovic2005environment}
Milenkovic, A., Milenkovic, M., Jovanov, E., Hite, D., Raskovic, D.: An
  environment for runtime power monitoring of wireless sensor network
  platforms. In: Proceedings of the Thirty-Seventh Southeastern Symposium on
  System Theory (SSST). pp. 406--410. IEEE (2005)

\bibitem{moon1997background}
Moon, U.K., Song, B.S.: Background digital calibration techniques for pipelined
  adcs. IEEE Transactions on Circuits and Systems II: Analog and Digital Signal
  Processing  44(2),  102--109 (1997)

\bibitem{potsch2017efficient}
P{\"o}tsch, A., Berger, A., Springer, A.: Efficient analysis of power
  consumption behaviour of embedded wireless iot systems. In: IEEE
  International on Instrumentation and Measurement Technology Conference
  (I2MTC). pp. 1--6. IEEE (2017)

\bibitem{AD5200}
\relax{Analog Devices Inc}: \relax{256-Position Digital Potentiometers}.
  \url{http://www.analog.com/media/en/technical-documentation/data-sheets/AD5200.pdf}
  (2012)

\bibitem{ADG1612}
\relax{Analog Devices Inc}: \relax{Quad SPST Switches}.
  \url{http://www.analog.com/media/en/technical-documentation/data-sheets/ADG1611_1612_1613.pdf}
  (2015)

\bibitem{Arduino}
\relax{Arduino Inc.}: \relax{ARDUINO MEGA 2560 REV3}.
  \url{https://store.arduino.cc/usa/arduino-mega-2560-rev3} (2017)

\bibitem{semiconductor2014official}
\relax{Freescale Semiconductor}: Official spi block guide v03. 06 (2014)

\bibitem{ieeeieee}
\relax{IEEE collaboration and others}: \relax{IEEE Standard for Terminology and
  Test Methods for Analog-To-Digital Converters}. IEEE Std pp. 1241--2000
  (2011)

\bibitem{agilent}
\relax{Keysight Technologies}: \relax{Keysight N6700 Modular Power System}.
  \url{https://www.keysight.com/en/pc-851482/n6700-modular-power-system?pm=SC}
  (2017)

\bibitem{MCP}
\relax{Microchip Inc.}: \relax{MCP3204/3208 2.7V 4-Channel/8-Channel 12-Bit A/D
  Converters}. \url{http://ww1.microchip.com/downloads/en/DeviceDoc/21298c.pdf}
  (2017)

\bibitem{k7510}
\relax{Tektronix, INC.}: \relax{DMM7510 7$\frac{1}{2}$-Digit Graphical Sampling
  Multimeter}.
  \url{https://www.tek.com/tektronix-and-keithley-digital-multimeter/dmm7510}
  (2017)

\bibitem{ina219}
\relax{Texas Instruments}: \relax{INA219 Zerø-Drift, Bidirectional
  Current/Power Monitor With I2C Interface}.
  \url{http://www.ti.com/lit/ds/symlink/ina219.pdf} (2015)

\bibitem{adc2015}
\relax{Texas Instruments}: \relax{Selecting an A/D Converter}.
  \url{http://www.ti.com/lit/an/sbaa004a/sbaa004a.pdf} (2015)

\bibitem{pot}
\relax{Vishay Spectrol}: \relax {Pot 10K$\Omega$}.
  \url{http://www.vishay.com/docs/57093/860.pdf} (2014)

\bibitem{suchanek2009adc}
Suchanek, P., Haasz, V., Slepicka, D.: Adc nonlinearity correction based on inl
  (n) approximations. In: IEEE International Workshop on Intelligent Data
  Acquisition and Advanced Computing Systems: Technology and Applications
  (IDAACS). pp. 137--140. IEEE (2009)

\bibitem{szewczyk2004analysis}
Szewczyk, R., Mainwaring, A., Polastre, J., Anderson, J., Culler, D.: An
  analysis of a large scale habitat monitoring application. In: Proceedings of
  the 2nd international conference on Embedded networked sensor systems. pp.
  214--226. ACM (2004)

\bibitem{taft20041}
Taft, R.C., Menkus, C.A., Tursi, M.R., Hidri, O., Pons, V.: A 1.8-v
  1.6-gsample/s 8-b self-calibrating folding adc with 7.26 enob at nyquist
  frequency. IEEE Journal of Solid-State Circuits  39(12),  2107--2115 (2004)

\bibitem{tsang2008background}
Tsang, C., Chiu, Y., Vanderhaegen, J., Hoyos, S., Chen, C., Brodersen, R.,
  Nikolic, B.: Background adc calibration in digital domain. In: IEEE Custom
  Integrated Circuits Conference (CICC). pp. 301--304. IEEE (2008)

\bibitem{zhou2013nemo}
Zhou, R., Xing, G.: Nemo: A high-fidelity noninvasive power meter system for
  wireless sensor networks. In: ACM/IEEE International Conference on
  Information Processing in Sensor Networks (IPSN). pp. 141--152. IEEE (2013)

\end{thebibliography}
\bibliographystyle{splncs03}

\end{document}